\documentstyle[aps,psfig,12pt]{revtex}

\begin{document}
\title{On thermostats and entropy production}
\author{Henk van Beijeren}
\address{Institute for Theoretical Physics, University of Utrecht, \\
Postbus 80006, Utrecht 3508 TA, The Netherlands}
\author{J. R.\ Dorfman}
\address{Institute for Physical Science and Technology, and\\
Department of Physics, \\
University of Maryland, College Park, Maryland,\\
20742, USA}
\date{\today}
\maketitle

\begin{abstract}
The connection between the rate of entropy production and
the rate of phase space contraction for thermostatted systems in
nonequilibrium steady states is discussed for a simple model of heat flow
in a Lorentz gas, previously described by Spohn and Lebowitz. It is easy
to show that for the model discussed here the two rates are not connected,
since the rate of entropy production is non-zero and positive, while the
overall rate of phase space contraction is zero. This is consistent with
conclusions reached by other workers. Fractal structures appear in the
phase space for this model and their properties are discussed. We conclude
with a discussion of the implications of this and related work for
understanding the role of chaotic dynamics and special initial conditions
for an explanation of the Second Law of Thermodynamics.
\end{abstract}

\section{Introduction}


There has been an intense interest over the
past several years in the properties of fluid systems subjected to
Gaussian 
thermostats\footnote{In fact Nos\'e-Hoover thermostats have been 
the preferred tool for doing simulations at a given temperature. These share
all properties discussed here for Gaussian thermostats\cite{HHP}.}.
These thermostats
are fictitious reversible force fields added to the equations of motion
for fluid systems, and were
introduced by people doing
molecular dynamics simulations on fluids in stationary nonequilibrium
systems in order to prevent the fluids from heating up
indefinitely\cite{gt,
hoover}.
These thermostatted systems have attracted the attention of workers in
statistical mechanics and in dynamical systems theory
for a
variety of reasons: Firstly, since phase space volumes are not conserved
in systems with Gaussian thermostats, and since the irreversible entropy
production in systems subjected to these type of thermostats can be
equated to the average rate of contraction of phase space volume on a
microscopic scale, it follows that the rate of entropy production is equal
to the negative of the sum of all Lyapunov exponents in the system
\cite{gt,
hoover,gtformalism}. Secondly, if, as often is the case, the
irreversible
entropy production may also be expressed in terms of transport
coefficients and thermodynamic driving forces, there is a direct link
between dynamical systems properties and nonequilibrium statistical
mechanics for such systems. Thirdly, a point of great interest is that the
phase space contraction induced by Gaussian thermostats in macroscopically
stationary nonequilibrium states causes these states to live on {\em fractal
attractors} that can be characterized by Sinai-Ruelle-Bowen (SRB)
measures. Gallavotti, Cohen and Ruelle\cite{gcr} have conjectured that
these may be the generic type of states needed to describe stationary
nonequilibrium states. Their Kaplan-Yorke dimension, which is related
directly to
sums of Lyapunov exponents, also measures the
irreversible entropy production as recently discussed by Evans {\it et }
{\it al.}\cite{ecsb}, in a way similar to that described by Gaspard for
Hamiltonian systems with escape\cite{gaspbook}. Finally,
Ruelle has argued\cite{Ruelle} that in a stationary state on a compact
subset of phase space, such as a shell of constant kinetic energy, the
average rate of phase space {\em contraction} cannot be negative, since
the phase space volume occupied by the state cannot expand without bounds.
Therefore the irreversible entropy production cannot be negative, which
would demonstrate the validity of the Second Law of thermodynamics for
Gaussian thermostated systems.

In order to avoid having
fictitious Gaussian thermostats
acting everywhere in the system, Chernov and Lebowitz\cite{cle} studied a
different class of thermostats that are active only at the boundaries of
the system. These thermostats act in such a way that no energy is
exchanged between thermostat and system, but the velocity distribution of
the particles reflected from the walls of the system is contracted in such
a way that a stationary shear flow results, giving rise to an irreversible
entropy production. In 
the infinite system limit the shear
induced irreversible entropy production could be equated again to the
average rate of phase space contraction produced by the thermostat, but
 other cases no such relation existed
for finite systems clear deviations from this equality were found. Obviously 
this could be due either
to irreversible entropy production in kinetic modes, which are always
present in boundary layers, or to a a complete breakdown of the relation
between phase space contraction rate and irreversible entropy production.
Related results have also been obtained by Klages and
coworkers\cite{klages}. These authors have constructed a variety of
thermostats which act only during collisions of particles with fixed
scatterers or walls, and also lead to nonequilibrium steady states, with
phase space contraction onto a lower dimensional attractor. In these
models, the entropy production may or may not be related to the rate of
phase space contraction, depending upon the precise nature of the
thermostat and the system.

Here we will show on the basis of a simple and fairly natural example that
indeed one can find thermostats for which the relation between the rate of
phase space contraction and the rate of entropy production breaks down,
and, in addition, we show one may have stationary nonequilibrium states
without an overall phase space contraction, as previously shown by
Gaspard\cite{gaspard}.
Our model consists of a
Lorentz gas enclosed between walls at different temperatures, a system for
which Spohn and Lebowitz proved long ago\cite{sl} that it satisfies
Fourier's law of heat conduction in the Grad limit, in which one sends the
size of the scatterers to zero while increasing their density in such a
way that the ratio between mean free path and system size remains
constant. The support of the stationary measure will have the same
dimension as the full phase space, yet it will exhibit strong fractal
features in the form of a {\em fractal repeller} 
located on the boundary
between two subsets of different energy, very similar to the fractal
boundaries studied by Gaspard in the framework of particle diffusion
\cite{gaspard}. The Lyapunov exponents of the system
will turn out to be simply related to those of the equilibrium system.
Finally we will comment on the consequences of our considerations for the
Second Law. We will argue that the assumptions used to obtain this law in
our context may be replaced by other, equally reasonable, assumptions
which lead to violations of the Second Law.

\section{The Lorentz gas}

The system
considered here is the usual Lorentz gas consisting of
fixed spherical scatterers located at random positions in a
$d$-dimensional space, and a light point particle moving between the
scatterers at constant speed and making specular, elastic collisions with
them.
We will assume that the scatterers cannot overlap each
other so as to avoid some problems with ergodicity and percolation that
may occur in the case of overlapping scatterers. Instead of this choice we
could also consider periodic billiards, with scatterers arranged on a
regular lattice (preferably without infinite horizons), or systems with
randomly oriented polyhedral scatterers instead of spherical ones. In the
latter case all Lyapunov exponents would be trivially zero, but, as in the
case of the 
periodic billiards, the transport properties would hardly
differ from those of the ordinary Lorentz gas\cite{windtree}. As
boundaries we choose two parallel flat walls, separated by a distance $L$.
The system may be infinite or periodically repeated in the directions
parallel to the walls. The collisions of the light particle with these
walls are also specular in direction, but the speed with which the
particle leaves a wall will always be uniformly distributed in the range
$v_{i}(1-\epsilon/2) \leq v \leq v_{
i}(1+\epsilon/2), i=1,2,$ where
$\epsilon <<1$, and $v_1=(d k_BT_1/m)^{1/2}$ is the average speed for one
wall and $v_2=(d k_BT_2/m)^{1/2}$, for the other\footnote{%
To be slightly
more realistic one could sample $v_1$ and $v_2$ from Maxwellian
distributions with temperatures $T_1$ respectively $T_2$, but for our
arguments this would
not make any difference.}. 
Eventually we have in mind the limiting case where $\epsilon$ tends to zero. 
Further, $m $ is the mass
of the light particle, $T$ is temperature, $k_B$ Boltzmann's constant and
$d$ the dimensionality. We suppose that the scatterers are distributed in
space with number density $n$, and that they have radius $a$. 

Spohn and
Lebowitz\cite{sl} have shown rigorously that in the Grad limit,
$n\rightarrow\infty,a\rightarrow 0,$ with $0< na^{d-1}<<1$, combined with
the limit of $L$ tending to infinity, this system satisfies Fourier's law
of heat conduction: the temperature profile becomes linear and the energy
current through the system becomes proportional to the temperature
gradient. In addition, the coefficient of heat conduction can be obtained
from the Lorentz-Boltzmann equation in this limit. One expects these
results to hold much more generally in the limit of large L, except that
for higher densities of scatterers (measured in dimensionless units $
\tilde{n}=n a^d$, with $\tilde{n}>0$) the transport coefficient will not
be 
following from the Lorentz-Boltzmann equation any more.

\section{Entropy production and dynamical properties}

It is easy
to find an expression for the irreversible entropy
production for these systems. As is usual for stationary thermostated
systems\cite{dvbphys} one has to argue that the irreversible entropy
production per unit time has to equal the rate of entropy 
flow
from the system into the thermostats. In the stationary state the average
energy current through the system has to have a constant value $j_e$, so
the average total entropy production per unit time $\sigma$ satisfies
\begin{equation}
\sigma=j_e(\frac 1 {T_2} -\frac 1 {T_1}),  \label{sigma}
\end{equation}
where we assume $T_1>T_2$. In the present model it follows
immediately that the energy current is directed from higher to lower
temperature, as every time a particle changes its velocity from $v_1$ to
$v_2$ it transfers an amount of energy $d k_B (T_1 - T_2)/2$ to the cold
reservoir, whereas in changes from $v_2$ to $v_1$ it extracts the same
amount of energy from the hot reservoir. It is not hard making explicit
estimates of $j_e$, especially for
small values of $\tilde{n}$, but
for our present purposes it suffices that this current is non-zero.
The phase space density of the particles traveling in the small range of
speeds about $v_{i}$ is easily seen to be proportional to
$v_{i}^{-(d+1)}$. This ensures, among other things, that the number of
particles incident on a small area of the wall per unit time is equal to
the number of particles leaving the small area per unit time. Furthermore,
under the action of the thermostats as defined above there is no {\em net}
contraction of phase space density. Indeed on a collision reducing $v_1$
to $v_2$ the phase space density at the coordinates of the collision is
multiplied by a factor $(v_1/v_2)^{(d+1)}$, but on the next collision
bringing the speed back to $v_1$ this contraction is undone exactly and
therefore the average rate of
phase space contraction over a long time becomes
exactly zero. As a result the identity between average phase space
contraction rate and irreversible entropy production that holds for
Gaussian thermostats, is not valid in the present case. In addition, the
presence of a stationary heat current influences the Lyapunov exponents in
a trivial way only. The crucial observation here is to note that the
trajectories of the light particle are exactly the same as in equilibrium,
only the parts of it that originate from the first wall are traversed at
speed $v_1$ and those originating from the second wall with speed $v_2$.
As a result the total time required for traversing a long trajectory is
multiplied by a factor of $(1/v_1 +1/v_2)/ (2/v)$ compared to the time
needed by a particle with constant speed $v$. Hence all the Lyapunov
exponents are multiplied by the inverse of this factor. The stationary
measure on phase space is a measure with 
one constant density, proportional to
$(1/v_1)^{(d+1)}$ and restricted to the velocity range $v_{1}(1-\epsilon/2) 
\leq v \leq v_{
1}(1+\epsilon/2)$, for points that can be traced back to a last collision
with the first wall, and another constant density,
proportional to $(1/v_2)^{(d+1)}$ and confined to the velocity range 
$v_{2}(1-\epsilon/2) \leq v \leq v_{
2}(1+\epsilon/2)$, for points coming from a last
collision with the other wall. It is completely analogous to the measures
considered by Gaspard to describe systems with a stationary diffusion
current\cite{gaspard}.

In spite of the absence of phase space contraction and the simplicity of
the Lyapunov spectrum these stationary measures have strong fractal
properties. Due 
to the way the model is constructed, the phase space
trajectories for the moving particles are located on two thin energy
shells about energies $mv_i^2/2, i=1,2$. In the stationary state, the
trajectories on each shell are those whose last collision with a wall led
to a value of the energy in the range of the shell. The trajectory of the
particle may, of course, take place on both shells. To better visualize
this phase space structure, it is helpful to rescale each of the energy
shells so that they coincide. Under these circumstances, one may visualize
the stationary state phase space regions as consisting of two disjoint
subsets, separated by a boundary. Each of the subsets consists of points
whose motion, extrapolated back in time reaches a particular wall. The
boundary between these two subsets
consists of points in phase space from which, extrapolating
back in time, neither boundary is ever reached. An even smaller subset of
this consists of points from which the boundaries are never reached either
in the forward or in the backward time direction. This latter set is
called a repeller, and it coincides with the intersection of its stable
and unstable manifolds, that is, with the intersection of the set of
points which may have collided with one or the other walls in the past but
will not do so in the future (the stable manifold of the repeller), or
have not collided with a wall in the past but may do so in the future (the
unstable manifold of the repeller). Therefore, it is clear that the
boundary of the two sets under discussion here coincides with the unstable
manifold of the repeller.
\footnote{The fractal
repeller may contain subsets of trajectories with an environment
extrapolating back entirely to a single boundary. These will not appear in
the boundary set considered presently. However, we expect that typical
trajectories on the repeller will extend throughout the system and get
arbitrarily close to either boundary. Hence the missing part should be of
vanishing measure relative to the main body of the repeller.}

The full fractal repeller described above plays a central role in the {\em %
escape rate formalism} of Gaspard and Nicolis\cite{escape}.
In this
formalism the escape rate from a system with open boundaries is related on
the one hand to the diffusion coefficient, $D$, and on the other hand to
the sum of positive Lyapunov exponents, $\lambda^{+}_i$ \ and the
Kolmogorov-Sinai (KS) entropy, $h_{KS}$, on the repeller.
For a Lorentz gas in two dimensions, Gaspard has shown how to
express the 
partial information dimension, $d_u$, of the unstable manifold of the
repeller
to the diffusion coefficient of the moving particle in the fixed array
of scatterers, and to the Lyapunov exponents and KS entropy\cite{gaspbook}.
In our
case, this fractal dimension is given by
\begin{eqnarray}
d_u & = & 
\frac{h_{KS}}{\lambda^{+}},  \nonumber \\
& = & 1-\frac{D\pi^{2}}{\lambda^{+} L^{2}} + o(L^2).
\end{eqnarray}
Simple scaling arguments show that this dimension is independent of
the velocity of the particle on the unstable manifold, so that our
scaling of the two energy shells to make them coincide does not affect
the dimension of the boundary separating the two sets of phase points.


In the
present system the coefficient of heat conduction may also be related to
properties of the repeller. This follows from the work of Dorfman and
Gaspard who derived a relation between the coefficient of heat
conductivity and the dynamical properties of trajectories on
an appropriate repeller for heat conduction\cite{dogas}. This repeller is
obtained by looking at the
fluctuations of the Helfand moments associated with heat
conduction. The Helfand moments undergo a deterministic diffusion in
phase space. When the magnitude of the Helfand moment for a
given trajectory in phase space exceeds a certain
value, say $(d/2)k_BT|L/2|$, one considers that the system has ``escaped'' from 
a
region in phase space, and the escape-rate formalism may be applied. In
the case of the Lorentz gas
under
discussion here, the Helfand moments for heat flow and for diffusion are
essentially the
same, differing only 
by a constant factor, since the moving particle keeps a
constant energy 
 one collision with a wall to the next. 
until collIding with a wall. Thus 
the
dimensional properties of
the stable and unstable manifolds of the repellers are the same in
both cases, diffusion and heat conductivity.




\section{Discussion}

The simple example discussed in this paper shows that macroscopically
stationary nonequilibrium measures need not be phase space contracting and,
consequentially, the average rate of irreversible entropy production need
not be equal to the average rate of phase space contraction. In fact it
seems to us that the thermostat considered here is closer to physical
reality than the Gaussian thermostats that do give rise to the identities
mentioned above. The net effect of our thermostat may be described as
mimicking a transport of phase space density between the hot and the cold
reservoir. As a result of this the temperatures of these reservoirs will
approach each other (though very slowly if the reservoirs are really large)
and the coarse grained entropy, based on a coarse grained phase space
density, will decrease in the hot reservoir, but increase by a larger amount
in the cold reservoir. This scenario can be made very plausible by
probabilistic arguments, but is very hard to prove for almost any
Hamiltonian model. Both Gaussian thermostats and the thermostat considered
above can be considered as simplified models of real, Hamiltonian
thermostats, in which the expected properties have been built in from
scratch. It does not appear plausible though that in realistic systems
coupled to thermostats at different temperatures the microscopic phase space
density will keep increasing forever without bounds. 
In view of Liouville's theorem physical changes in phase space density always 
must be 
the result of exchanges of phase space density between the system and its 
thermostats. If the latter may be treated as stationary sources the distribution 
of phase space density in a stationary nonequilibrium state for the system has 
to be stationary as well. 
In our opinion 
the feature of steady phase space contraction really is an artifact of Gaussian 
thermostats. This even raises the
question whether there is a real use for 
these besides their handiness in
nonequilibrium steady state simulations, but there
certainly is. First of all
Gaussian thermostats can be used as dynamical probes of nonequilibrium
properties, precisely because of their satisfying the equality of entropy
production and phase space contraction rate. Further, there are strong
indications that at least for systems not too far from equilibrium, Gaussian
thermostats lead to behavior that is almost indistinguishable from that
brought about by more physical thermostats\cite{gt,sarman,thermos}. The
nonequilibrium features of stationary nonequilibrium measures show up
predominantly in their fractal features. The latter are much more visible
for systems with Gaussian thermostats, in which the stationary measure lives
on a fractal exclusively, than in the presumably more physical stationary
measures in which the fractal features only show up in the boundaries
between subsets with 
different densities. Therefore systems with Gaussian
thermostats may be more suitable in the end for a dynamical study of
nonequilibrium states. At this moment it is unclear whether this really is
the case and in addition the relations between systems with different types
of thermostats still largely have to be established.


We now make some remarks on the implications of thermostating mechanisms
for the Second Law of thermodynamics. We indicated already that the
generation of an average energy current from the hot to the cold reservoir
has been built into our thermostat by hand. To make this point somewhat
clearer, let us consider a somewhat more physical thermostat. As illustrated
in Figure 1 it consists of a gas of particles that do not see the Lorentz
gas walls, but instead are confined by a wall somewhat inside the Lorentz
gas region. The latter wall in turn is invisible to the light Lorentz gas
particle. In
the narrow strip between these two walls the Lorentz gas particle may
collide with the thermostat particles and so, if these collisions are
sufficiently frequent, the light particle will leave the strip with a
kinetic energy that on average is proportional to the temperature of the
thermostat.
\begin{figure}[b]
\centerline{\psfig{figure=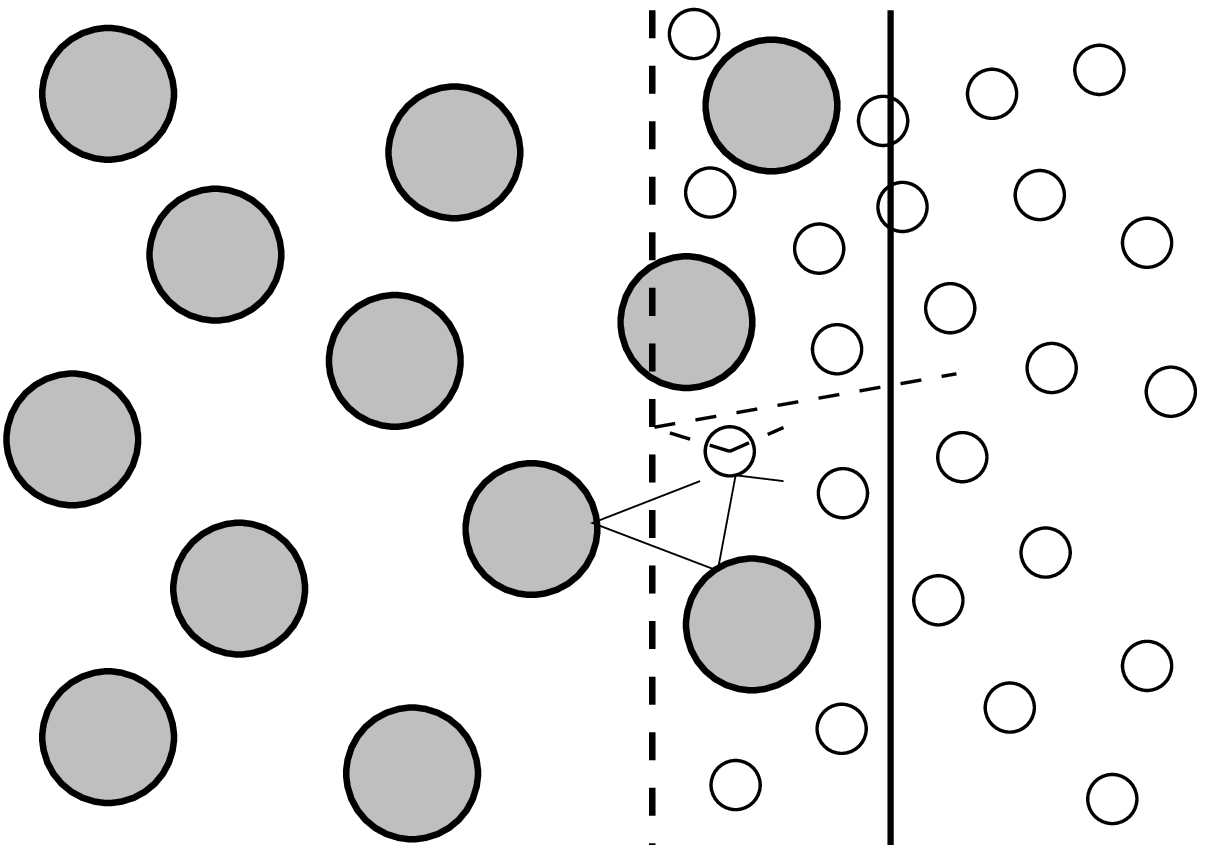,angle=-90,height=6.0cm}}
\caption{{\bf A more physical thermostat.} This figure shows a small part of 
a boundary region of a Lorentz gas with a more pysical thermostat. The large 
circles are fixed
scatterers and the small circles are moving thermostat particles. The solid
line indicates the wall of the Lorentz gas, which is impenetrable to the
light Lorentz gas particle but invisible to the thermostat particles. The
dashed line indicates the wall of the thermostat, which is invisible to the
Lorentz gas particles. The solid and dashed broken line describe part of a
trajectory of the light Lorentz gas particle and a thermostat particle
respectively.}
\end{figure}
The combined action of two of these thermostats on either side,
with temperatures $T_1$ and $T_2$, will be basically the same as that
of our simple model thermostats. For this to be true, however, we
implicitly have to break time reversal symmetry by assuming that the
velocity distributions of the Lorentz particle and of the thermostat
particles {\em before} their collision are mostly uncorrelated, in other
words
we
have to assume Boltzmann's Stoszzahlansatz or some generalization of
it. 
Instead, we could have required such property to hold {\em after} the
collision
\cite{cober}.
This would correspond to the macroscopic arrow of time running
in the
opposite direction. Of course, we as observers would not notice the
difference. Only an outside observer could notice our inversion of the
notions of past and future. But, as remarked by e.g. Schulman\cite{schulman},
more exotic possibilities cannot be excluded. The arrow of time could run
one way in parts of the system and the other way in other parts. Also, on
the basis of known microscopic equations of motion one cannot exclude the
possibility that at some time the direction of the macroscopic arrow of time
will turn around, perhaps through growth of the area in which it runs the
opposite way. To explain our obvious observation of a well defined direction
of time, reversible microscopic equations simply do not suffice.
Apparently we do need
the additional fact that at some point in time our universe was brought into
a state of extremely low entropy, so we observe an arrow of time moving away
quite definitely from that point.
And in
spite of any lack of evidence we cannot exclude the possibility that the
temporal boundary conditions to the `equations of motion of our universe'
are not all located in the past, which then might lead indeed to eventual
inversions of the arrow of time.
In view of this it
seems clear that proofs of the Second Law will have to require
additional information about the properties and the history of our universe
and cannot be based 
exclusively upon the mixing or chaotic 
properties of the microscopic equations of motion, or
on properties of thermostats in stationary nonequilibrium systems. In
less cosmic language, we might say that some assumptions about the
absence of initial correlations in our systems will have to be made in
order to demonstrate the validity of the Second Law, no 
matter how
chaotic the dynamics of our system. Along these lines, a much
more modest goal, which is still very hard to attain, is to show that
isolated systems showing sufficient lack of correlations initially, will
indeed satisfy the Second Law. Here perhaps thermostats can be of some use,
but in order to be credible, they would have to be of Hamiltonian type .

\section{Acknowledgements}

For both of us Joel Lebowitz has been a dear friend for over thirty five
years. It is a great pleasure being able to contribute to this very special
Festschrift. We congratulate Joel on his seventieth birthday and express
our hope he will continue for many more years being an inspiring example for
all statistical physicists
in all his scientific and humanitarian efforts.
We thank Rainer Klages for some valuable remarks on the manuscript. 
JRD wishes to thank FOM for support of visits to the University
of Utrecht, and the National Science Foundation (USA) for support
under grants PHY-96-00428 and PHY-98-20824. HvB acknowledges support by FOM,
SMC and by the NWO Priority Program Non-Linear Systems, which are
financially supported by the "Nederlandse Organisatie voor Wetenschappelijk
Onderzoek (NWO)".

\end{document}